\documentclass{article}
\usepackage[T1]{fontenc} 
\usepackage[utf8]{inputenc} 
\usepackage{ismir,amsmath,cite,url}
\usepackage{graphicx}
\usepackage{color}
\usepackage{todonotes}
\usepackage{multirow}
\usepackage{subcaption}
\usepackage{mathtools}
\usepackage{amssymb}
\usepackage{amsthm}
\usepackage{rotating}
\usepackage{algorithm}
\usepackage{algpseudocode}
\DeclarePairedDelimiterX{\infdivx}[2]{(}{)}{%
  #1\;\delimsize\|\;#2%
}

\DeclarePairedDelimiter{\norm}{\lVert}{\rVert}

\title{Evaluating Recommender System Algorithms for Generating Local Music Playlists}

\threeauthors
  {Daniel Akimchuk} {Ithaca College \\ {\tt dakimchuk@ithaca.edu}}
  {Timothy Clerico} {Ithaca College \\ {\tt tclerico@ithaca.edu }}
  {Douglas Turnbull} {Ithaca College \\ {\tt dturnbull@ithaca.edu}}

\begin{document}

\maketitle
\begin{abstract}
We explore the task of \emph{local} music recommendation: provide listeners with personalized playlists of relevant tracks by artists who play  most of their live events within a small geographic area.  Most local artists tend to be obscure, long-tail artists and generally have little or no available user preference data associated with them. This creates a cold-start problem for collaborative filtering-based recommendation algorithms that depend on large amounts of such information to make accurate recommendations.  In this paper, we compare the performance of three standard recommender system algorithms (Item-Item Neighborhood (IIN), Alternating Least Squares for Implicit Feedback (ALS), and Bayesian Personalized Ranking (BPR)) on the task of local music recommendation using the Million Playlist Dataset. To do this, we modify the standard evaluation procedure such that the algorithms only rank tracks by local artists for each of the eight different cities. Despite the fact that techniques based on matrix factorization (ALS, BPR) typically perform best on large recommendation tasks, we find that the neighborhood-based approach (IIN) performs best for long-tail local music recommendation. 


\end{abstract}
\section{Introduction}\label{sec:intro}

If you were to move to a new city and wanted to check out the local music scene, how would you get started? You might ask an expert, such as an employee at a local music store or a barista at a local coffee shop, but they are likely to give you incomplete or biased recommendations based on their own personal experiences and interests. You might also pick up the arts section of the local newspaper or go online to find a community events notice board. Either way, you would be faced with a long list of music events, each of which would only provide a small amount of contextual information such as artist names and perhaps a few genre labels.

Music recommender systems \cite{schedl2018current} have the potential to offer an alternative to these more traditional methods of exploring the local music scene. However, the most popular music streaming services (e.g. Spotify, Pandora, Apple Music, Deezer) offer little, if any, support of music discovery based on geographic region. For example, if a user wants to find music from a specific location on Spotify, they would have to use the generic text-based search functionality and then dig through playlists with that location's name in the playlist title or description. Often, even if such playlists exist, they are outdated, not personalized to match the user's interests, and may not be relevant due to a variety of factors (e.g. cities with common names, playlists with non-local music, etc.).    

By contrast, music event recommendation services like BandsInTown\footnote{https://www.bandsintown.com} and SongKick\footnote{https://www.songkick.com/}  help users follow artists so that that the user can be notified when a favorite artist will be playing nearby. They also recommend upcoming events with artists who are similar to one or more of the artists that the user has selected to follow. These services have been successful in growing both the number of users and the number of artists and events covered by their service. For example, BandsInTown claims to have 38 million users and lists events for over 430,000 artists\footnote{According to https://en.wikipedia.org/wiki/Bandsintown on March 28, 2018.}. Event listings are added by aggregating information of ticket sellers (e.g. Ticketmaster\footnote{https://www.ticketmaster.com/}, TicketFly\footnote{https://www.ticketfly.com/}) and by managers and booking agents who have the ability to directly upload tour dates for their touring artists to these services.

While this coverage is impressive, a large percentage of the events found in local newspapers are not listed on these commercial music event recommendation services. Many talented artists play at small venues like neighborhood pubs, coffee shops, and DIY shows, and are often not represented by (pro-active, tech-savvy) managers. Yet many music fans enjoy the intimacy of a small venue and a personal connection with local artists, and they may have a hard time discovering these events. 

Our long-term goal is to create a locally-focused music recommender system that (1) helps users create personalized playlists that feature relevant music by local artists, and (2) provides users with personalized music event recommendations. A core component of this system is to explore how existing recommender system algorithms perform to the task of local music recommendation. Here we consider a \emph{local} artist to be an artist or band who resides in and/or plays the majority of their live music events in a small geographic region such as a city (e.g. Liverpool, Seattle) or a neighborhood within a larger city (e.g. Haight-Ashbury in San Francisco, Gràcia in Barcelona).

\begin{figure}
\centering
\includegraphics[width=9cm]{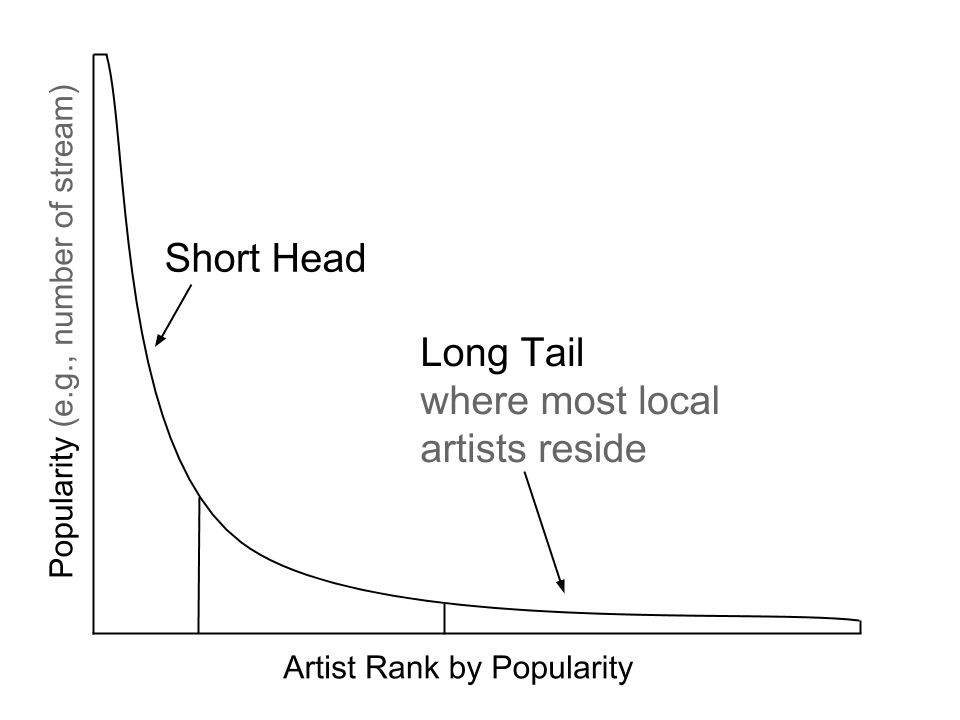}
\caption{Long-Tail Distribution for Music Consumption}
\label{fig:longtail}
\end{figure}

\subsection{Long-tail Recommendation \& Popularity Bias}
Local music recommendation can be considered a special case of the \emph{long-tail} music recommendation problem \cite{levy2010music, celma2010music, anderson2006long} since most local artists are relatively obscure outside of their home cities. The long-tail metaphor \cite{anderson2006long} comes from the idea that if we order each artist by popularity and plot how many times their music is consumed (i.e. purchased/downloaded/streamed) we would see a rapid drop off (i.e. power-law distribution) such that that a very small fraction of the artists (in the short-head) would receive the majority of the consumption while the overwhelming majority of artists (in the long-tail) receive little or no attention (see Figure \ref{fig:longtail}).

Recommender systems are known to suffer from \emph{popularity bias} \cite{bellogin2017statistical, celma2008hits}: popular artists are recommended often while obscure, long-tail, artists are rarely recommended, if at all. This creates a feedback loop in which ``the rich get richer'' and prevents local artists from being discovered by potential fans. Popularity bias is manifested in (commercial) recommender systems due to a combination of conceptual and technical reasons. First, listeners tend to prefer familiar music \cite{north1995subjective, hunter2011interactive}, so it is safer for a music streaming service to recommend popular songs or artists that are more likely to be known to the user. Second, recommender systems that use aggregated user preference data, known as collaborative filtering (CF) systems, suffer the cold-start problem \cite{schedl2018current, turnbull2008five}: little or no historical user preference data exists for new or obscure artists. As a result, a CF-based recommender system cannot recommend these artists with sufficient confidence.   

In this paper, we explore how existing recommender system algorithms perform on the task of local music recommendation. We formulate this problem as a modification of the \emph{automatic playlist continuation task} \cite{schedl2018current} that was the focus of the 2018 ACM RecSys Challenge\footnote{https://recsys-challenge.spotify.com/} \cite{chen2018recsys}. Specifically, we evaluate how accurately different recommender systems predict additional tracks for existing playlists, but we limit the additional tracks to be those by artists who are associated with a given city or neighborhood. We consider this formulation to be a case study in how different recommender system algorithms perform at the task of long-tail recommendation.

\section{Recommender System Algorithms}\label{sec:recsys}
\begin{figure}
\centering
\includegraphics[width=8cm]{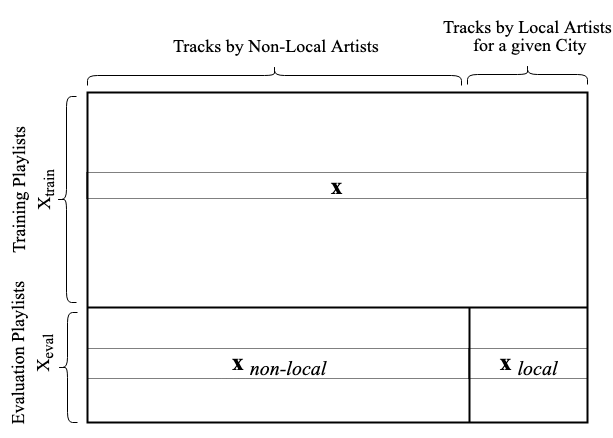}
\caption{Playlist-Track Matrix: We train each recommender system model with the training playlists and evaluate them using the evaluation playlists. We partition each evaluation playlist and use $\mathbf{x}_{non-local}$ as input to the model. The model then scores each local track $\hat{\mathbf{x}}_{local}$ (not shown) and evaluate performance by comparing $\hat{\mathbf{x}}_{local}$ to the ground truth $\mathbf{x}_{local}$.}
\label{fig:eval}
\end{figure}

In this section we describe three common recommendation algorithms: Item-Item Neighborhood (IIN) Recommendation,  Alternating Least Squares (ALS) for Implicit Feedback, and Bayesian Personalized Ranking (BPR). Our main data structure is a Playlist-Track matrix which is akin to a User-Item matrix in standard CF research.

Each of the algorithms described takes as input this $m\times n$ matrix $X$ like the one shown in Figure \ref{fig:eval}. The element of this matrix in the $p$-th row and $t$-th column, denoted $x^{(p)}_t$, reflects the rating of the track $t$ in playlist $p$. We consider our data to represent \emph{implicit} feedback where the value of $x^{(p)}_{t}$ is 1 if track $t$ is found in playlist $p$ and 0 otherwise. 

\subsection{Item-Item Neighborhood Model (IIN)}\label{sec:iin}

Neighborhood models are traditional collaborative filtering methods that make recommendations based on the similarities of playlists and/or tracks \cite{volkovs2018two, iin}. Item-Item Neighborhood (IIN) models focus specifically on the similarity of different tracks. They function under the assumption that if a track is similar to the tracks already associated with a playlist, then it is likely to be a successful recommendation. 

Given playlist $p$ and track $t$, the Item-Item similarity score, $s^{(p)}_t$, can be calculated via $$s^{(p)}_{t} = \sum_{t'\in \mathcal N(p)} \frac{\mathbf{x}_{t}\mathbf{x}_{t'}}{\norm{
\mathbf{x}_{t}}\norm{\mathbf{x}_{t'}}},$$ where $\mathcal N(p)$ is the set of nonzero tracks in playlist $p$, and $\mathbf{x}_{t}$ is the $t$-th column of $X$, containing the ratings from each playlist for track $t$.
To recommend tracks for playlist $p$, $s^{(p)}_{t}$ is calculated for every track $t$ and sorted from greatest to least. 


\subsection{Alternating Least Squares for Implicit Feedback (ALS)}
\label{als}

Weighted Regularized Matrix Factorization (WRMF) optimized by Alternating Least Squares (ALS) \cite{als} is one of the most highly-cited and successful recommender system models. For example, it was the model used to win the Netflix Prize \cite{koren2009matrix} in 2009 and was an integral component of the system that recently won the 2018 ACM RecSys Challenge that focused on \emph{music playlist continuation} \cite{volkovs2018two}. The goal of this algorithm is to map playlists and tracks into a common latent factor space in which they can be compared.

To address the case where $x^{(p)}_{t}$ can be a value other than 0 or 1, we define 
$$
r^{(p)}_{t} = \left\{
        \begin{array}{ll}
            1 & x^{(p)}_{t} > 0 \\
            0 & x^{(p)}_{t} = 0
        \end{array}
    \right.
.$$
In this case, it also proves helpful to define a confidence value $c^{(p)}_{t}$ for $r^{(p)}_{t}$. While there are many options, Hu et al. \cite{als} suggest using $$c^{(p)}_{t} = 1 + \alpha x^{(p)}_{t},$$ with hyperparameter $\alpha$. 

The latent factors $\mathbf{f}^{(p)}$ for each playlist $p$ and $\mathbf{f}_t$ for each track $t$, both elements of $\mathbb{R}^f$, are to be computed with the goal that $r^{(p)}_{t} = \mathbf{f}_{t}^T\mathbf{f}^{(p)}.$ This will be done by minimizing the cost function $$\min\sum_{p,t}c^{(p)}_{t}\left(r^{(p)}_{t}-\mathbf{f}_{t}^T\mathbf{f}^{(p)}\right)^2 + \lambda\left(\sum_p\norm{\mathbf{f}^{(p)}}^2 + \sum_t\norm{\mathbf{f}_t}^2\right).$$ Note the $\lambda$ term is used for regularization.

This sum has $mn$ terms, which makes it computationally impractical to use traditional cost minimization, so instead we repeatedly recompute the playlist factor $\mathbf{f}^{(p)}$ and the track factor $\mathbf{f}_t$. First, to recompute the playlist factors, we define an $n\times f$ matrix $Y_f$. Each row of this matrix is the track factor for a given track. We also define an $n\times n$ diagonal matrix $C^{(p)}$ for each playlist $p$ such that $$C^{(p)} = \begin{bmatrix} 
c^{(p)}_{1} & 0 & ... & 0 \\
0         & c^{(p)}_{2} & ... & 0 \\
\vdots & & \ddots & \\
0 & 0 & ... & c^{(p)}_{n}
\end{bmatrix}.$$ With $\mathbf{r}^{(p)}$ being the $n$-dimensional vector of all $r^{(p)}_t$, we minimize the cost function with $$\mathbf{f}^{(p)} = \left(Y_f^TC^{(p)}Y_f + \lambda I\right)^{-1}Y_f^TC^{(p)}\mathbf{r}^{(p)}.$$

In a similar manner, at the same time we recompute the track factors by defining an $m\times f$ matrix $X_f$ with each row being the playlist factor for a given playlist. We also define a similar diagonal matrix, this time of size $m \times m$, $C_t$ with $$C_t = \begin{bmatrix} 
c^{(1)}_t & 0 & ... & 0 \\
0         & c^{(2)}_t & ... & 0 \\
\vdots & & \ddots & \\
0 & 0 & ... & c^{(m)}_t
\end{bmatrix}.$$ With $\mathbf{r}_t$ being the $m$-dimensional vector of all $r^{(p)}_t$, we minimize the cost function with $$\mathbf{f}_t = \left(X_f^TC_tX_f + \lambda I\right)^{-1}X_f^TC_t\mathbf{r}_t.$$ This is repeated until $\mathbf{f}^{(p)}$ and $\mathbf{f}_t$ stabilize.

We predict the preference of playlist $p$ for track $t$ via $$\hat{r}^{(p)}_t = \mathbf{f}_t^T\mathbf{f}^{(p)}.$$

\subsection{Bayesian Personalized Ranking}
\label{bpr}
Traditional methods of training recommendation algorithms assume all non-ranked tracks by a playlist to have a rank of 0. This implies that the ``perfect" algorithm would give these tracks a rank of a 0. However, as we want to rank these zero-valued tracks, this isn't our desired output and is adjusted for using regularization to avoid this overfitting. As described by Rendle et al. \cite{bpr}, Bayesian Personalized Ranking attempts to address this issue without the need for regularization. It defines a new optimization criterion on which to train a model.

BPR functions under the assumption that any track that is in a playlist (any track that has a nonzero rating in $\mathbf{x}^{(p)}$) is preferred by that playlist over a track that is not in the playlist. To formalize this, we define two sets: the set of playlists $P$ and the set of tracks $T$, and also the set $$D = \left\{(p,\,t,\,t')\hspace{5pt}\middle|\hspace{5pt}\text{playlist } p \text{ prefers track } t \text{ over } t'\right\}.$$ If $(p,\,t,\,t')\in D$, then $t >_p t'$, where $>_p$ is the preference structure for playlist $p$.

Let $\Theta$ be the parameters of the underlying learning algorithm (the implementation used in this experiment utilizes matrix factorization). By Bayes' Law, we know the probability of $\Theta$ being the correct parameter vector given playlist $p$'s preference structure $$\text{p}(\Theta\hspace{1pt}|\hspace{1pt}>_p) \propto \text{p}(>_p\hspace{1pt}|\hspace{1pt}\Theta)\text{p}(\Theta).$$ It also follows that $$\prod_{p\in P} \text{p}(>_p\hspace{1pt}|\hspace{1pt}\Theta) = \prod_{(p,t,t')\in D}\text{p}(t >_p t'\hspace{1pt}|\hspace{1pt}\Theta).$$

Using the underlying learning algorithm, the predicted relationship between tracks $t$ and $t'$ for playlist $p$ using the parameters $\Theta$, referred to as $\hat{x}^{(p)}_{tt'}(\Theta)$, is to be calculated. We assign $$\text{p}(t >_p t'\hspace{1pt}|\hspace{1pt}\Theta) = \sigma\left(\hat{x}^{(p)}_{tt'}(\Theta)\right),$$ where $\sigma$ is the sigmoid function.

The prior probability of $\Theta$, $\text{p}(\Theta)$, is a normal distribution with zero mean and variance-covariance matrix $\Sigma_\Theta$. As suggested in \cite{bpr}, we use $\Sigma_\Theta = \lambda_\Theta I$, where $\lambda_\Theta$ is the vector of regularization parameters for the underlying learning algorithm and $I$ is the identity matrix.

Using these identities, the optimization criterion to maximize $\ln \text{p}(\Theta\hspace{1pt}|\hspace{1pt}>_p)$ can be written as the calculable $$\sum_{(p,t,t')\in D} \ln \sigma\left(\hat{x}^{(p)}_{tt'}\right) - \lambda_\Theta\norm{\Theta}^2.$$ For a more detailed derivation of this optimization criterion, see \cite{bpr}. This is maximized using any optimization algorithm, such as gradient descent, and the resulting $\Theta$ parameter vector is used with the underlying learning algorithm.

\section{Local Music Data}\label{sec:data}

Our first task is to identify a set of local artists for a given city. For the paper, we consider a local artist to be an artist that performs the large majority of their live events close to or within a single city. We collected artist event information from both Ticketfly\footnote{http://www.ticketfly.com/} and Facebook\footnote{https://www.facebook.com/}. Ticketfly provides information about large and mid-sized events while Facebook provides information about smaller niche events that were not listed on Ticketfly.  We were able to collect 22,246 unique events at over 3,500 different venues for over 145,000 artists for a span of 3 months\footnote{All event data was collected in February 2019.}. Of these events, 17,976 events come from Ticketfly and 8,447 from Facebook, with an overlap of 4,177 events between the two sites. We associate an artist as being local to a city if at least 80\% of their events were within a 10-mile radius of city center and they were associated with at least 2 events.

For this study, we selected a geographically diverse set of eight cities within the United States. For each city, we create the list of local artists from our music event data and collect the set of tracks by these artists. Finally, we identify all of the playlists from the the Million Playlist Dataset \cite{chen2018recsys} that contains one or more of these tracks.  A list of the cities as well as summary statistics about each city can be found in Table \ref{table::summary}. 

We note that most of the local artists in our study are obscure long-tail artists and tend to have between a few hundreds to a few thousand of monthly listeners on Spotify. This is also reflected in the fact that the \emph{sparsity} (i.e. percent of zeros) for the columns in the playlist-track matrix $X$ associated with the tracks by the local artists is extremely sparse (average of 99.9990\% sparse.) This make the task of local music recommendation particularly challenging when we consider that the overall sparsity of $X$ is 99.9971\%. Put another way, the overall density (percent of non-zero ratings) is about 3 times more dense than the density for local (long-tail) artists.

 \begin{table*}
 \resizebox{\textwidth}{!}{

\begin{tabular}{| c | c | c | c | c | c | c | c | c | c |}
\hline
&Atlanta & Berkeley & Boulder & Brooklyn & Chicago & Los Angeles & Nashville & Philadelphia & Average \\
\hline
Local Playlists & 221 & 1022 & 110 & 886 & 84 & 375 & 98 & 276 &    384.0\\
\hline
Local Artists & 15 & 41 & 6 & 123 & 12 & 36 & 16 & 39 &     36.0\\
\hline
Median Monthly Listeners & 4,311 & 1,677 & 39,609 & 2,017 & 84 & 2,370 & 1,244 & 72 & 6,423\\
\hline
Local Tracks & 388 & 2023 & 237 & 1468 & 260 & 556 & 140 & 519 &    698.9\\
\hline
Sparsity &     99.9991\% &     99.9990\% &     99.9983\% &     99.9989\% &    99.9997\% &     99.9978\% &     99.9991\% &    99.9997\% &     99.9990\% \\
\hline
\end{tabular}
}
\caption{Summary statistics for each city evaluated.}
\label{table::summary}
\end{table*}

\begin{table*}
\resizebox{\textwidth}{!}{
\begin{tabular}{| c | c | c | c | c | c | c | c | c | c | c |}
\hline
\multicolumn{11}{|l|}{Tracks} \\
\hline
& & Atlanta & Berkeley & Boulder & Brooklyn & Chicago & Los Angeles &  Nashville & Philadelphia & Average \\
\hline
\multicolumn{2}{|l|}{Local Tracks} & 388 & 2023 & 237  & 1468 & 260 & 556 & 140 & 519 & 698.875\\
\hline
\multirow{5}{*}{\rotatebox[origin=c]{90}{NDCG}} & Item-Item & \textbf{0.335 (0.041)} & \textbf{0.295 (0.033)} &  \textbf{0.268 (0.052)} &\textbf{0.324 (0.056)} & 0.199 (0.044) & \textbf{0.339 (0.074)} & \textbf{0.447 (0.103)} & \textbf{0.355 (0.049)} & \textbf{0.324}\\
\cline{2-11}
& ALS & 0.065 (0.065) & 0.046 (0.046) &  0.066 (0.066) &0.042 (0.042) & 0.057 (0.057) & 0.043 (0.043) & 0.086 (0.086) & 0.036 (0.036) & 0.055\\
\cline{2-11}
& BPR & 0.036 (0.036) & 0.025 (0.025) &  0.036 (0.036) &0.026 (0.026) & 0.048 (0.048) & 0.030 (0.030) & 0.038 (0.038) & 0.031 (0.031) & 0.034\\
\cline{2-11}
& Random & 0.177 (0.005) & 0.133 (0.002) &  0.207 (0.008) &0.131 (0.001) & 0.211 (0.013) & 0.161 (0.002) & 0.208 (0.013) & 0.166 (0.005) & 0.174\\
\cline{2-11}
& Popular & 0.225 (0.010) & 0.161 (0.002) &  0.248 (0.026) &0.159 (0.003) & \textbf{0.262 (0.019)} & 0.179 (0.004) & 0.255 (0.011) & 0.182 (0.008) & 0.209\\
\cline{2-11}
\hline
\multirow{5}{*}{\rotatebox[origin=c]{90}{RPrec}} & Item-Item & \textbf{0.100 (0.017)} & \textbf{0.077 (0.010)} &  \textbf{0.099 (0.028)} &\textbf{0.114 (0.017)} & 0.046 (0.020) & \textbf{0.091 (0.020)} & \textbf{0.148 (0.054)} & \textbf{0.152 (0.028)} & \textbf{0.103}\\
\cline{2-11}
& ALS & 0.026 (0.026) & 0.012 (0.012) &  0.023 (0.023) &0.011 (0.011) & 0.005 (0.005) & 0.008 (0.008) & 0.022 (0.022) & 0.001 (0.001) & 0.014\\
\cline{2-11}
& BPR & 0.004 (0.004) & 0.000 (0.000) &  0.000 (0.000) &0.001 (0.001) & 0.005 (0.005) & 0.000 (0.000) & 0.004 (0.004) & 0.000 (0.000) & 0.002\\
\cline{2-11}
& Random & 0.008 (0.003) & 0.002 (0.000) &  0.016 (0.005) &0.001 (0.000) & 0.015 (0.007) & 0.009 (0.003) & 0.015 (0.014) & 0.006 (0.002) & 0.009\\
\cline{2-11}
& Popular & 0.035 (0.012) & 0.018 (0.001) &  0.046 (0.020) &0.022 (0.003) & \textbf{0.058 (0.018)} & 0.013 (0.004) & 0.042 (0.012) & 0.014 (0.007) & 0.031\\
\cline{2-11}
\hline
\multirow{5}{*}{\rotatebox[origin=c]{90}{Prec@1}} & Item-Item & \textbf{0.117 (0.039)} & \textbf{0.095 (0.032)} &  \textbf{0.094 (0.028)} &\textbf{0.128 (0.010)} & 0.032 (0.020) & \textbf{0.090 (0.004)} & \textbf{0.120 (0.046)} & \textbf{0.187 (0.014)} & \textbf{0.108}\\
\cline{2-11}
& ALS & 0.036 (0.036) & 0.020 (0.020) &  0.027 (0.027) &0.015 (0.015) & 0.000 (0.000) & 0.016 (0.016) & 0.032 (0.032) & 0.004 (0.004) & 0.019\\
\cline{2-11}
& BPR & 0.005 (0.005) & 0.000 (0.000) &  0.000 (0.000) &0.001 (0.001) & 0.000 (0.000) & 0.000 (0.000) & 0.000 (0.000) & 0.000 (0.000) & 0.001\\
\cline{2-11}
& Random & 0.000 (0.000) & 0.001 (0.001) &  0.027 (0.018) &0.005 (0.003) & 0.012 (0.012) & 0.011 (0.008) & 0.011 (0.011) & 0.007 (0.004) & 0.009\\
\cline{2-11}
& Popular & 0.023 (0.012) & 0.015 (0.003) &  0.073 (0.031) &0.034 (0.006) & \textbf{0.035 (0.024)} & 0.019 (0.007) & 0.010 (0.010) & 0.018 (0.010) & 0.028\\
\cline{2-11}
\hline
\end{tabular}
}
\caption{Evaluation metrics at the track-level for each algorithm.}
\label{table::track}
\end{table*}

\begin{table*}
\resizebox{\textwidth}{!}{
\begin{tabular}{| c | c | c | c | c | c | c | c | c | c | c |}
\hline
\multicolumn{11}{|l|}{Artists} \\
\hline
& & Atlanta & Berkeley &  Boulder &Brooklyn & Chicago & Los Angeles &  Nashville & Philadelphia & Average \\
\hline
\multicolumn{2}{|l|}{Local Artists} & 15 & 41 &  6 &123 & 12 & 36 & 16 & 39 & 36 \\
\hline
\multirow{5}{*}{\rotatebox[origin=c]{90}{NDCG}} & Item-Item & 0.751 (0.076) & \textbf{0.725 (0.059)} &  0.833 (0.086) &\textbf{0.604 (0.088)} & 0.652 (0.057) & \textbf{0.732 (0.128)} & 0.713 (0.108) & \textbf{0.686 (0.065)} & \textbf{0.712}\\
\cline{2-11}
& ALS &      \textbf{0.793 (0.020)} & 0.691 (0.005) &  \textbf{0.928 (0.019)} &0.453 (0.009) & \textbf{0.824} (0.048) & 0.617 (0.025) & \textbf{0.732 (0.025)} & 0.601 (0.019) & 0.705\\
\cline{2-11}
& BPR &      0.503 (0.006) & 0.359 (0.017) &  0.699 (0.032) &0.240 (0.006) & 0.732 (0.028) & 0.407 (0.012) & 0.570 (0.030) & 0.328 (0.016) & 0.480\\
\cline{2-11}
& Random &      0.580 (0.011) & 0.415 (0.009) &  0.690 (0.021) &0.303 (0.007) & 0.744 (0.022) & 0.443 (0.007) & 0.547 (0.032) & 0.443 (0.012) & 0.521\\
\cline{2-11}
& Popular &      0.561 (0.018) & 0.421 (0.007) &  0.811 (0.042) &0.386 (0.005)& 0.759 (0.028) & 0.470 (0.011) & 0.577 (0.015) & 0.458 (0.014) & 0.555\\
\cline{2-11}
\hline
\multirow{5}{*}{\rotatebox[origin=c]{90}{RPrec}} & Item-Item &      \textbf{0.436 (0.037)} & \textbf{0.377 (0.032)} &       \textbf{0.543 (0.048)} &\textbf{0.319 (0.050)} & \textbf{0.340 (0.041)} & \textbf{0.431 (0.079)} & \textbf{0.543 (0.074)} & \textbf{0.427 (0.058)} & \textbf{0.427}\\
\cline{2-11}
& ALS &      0.284 (0.010) & 0.247 (0.007)&  0.405 (0.043) &0.127 (0.010)& 0.323 (0.067) & 0.164 (0.022) & 0.364 (0.032) & 0.264 (0.022) & 0.272\\
\cline{2-11}
& BPR &      0.089 (0.015) & 0.043 (0.009) &  0.186 (0.050) &0.017 (0.005)& 0.263 (0.038) & 0.050 (0.007) & 0.191 (0.042) & 0.028 (0.008) & 0.108\\
\cline{2-11}
& Random &      0.090 (0.017) & 0.065 (0.005) &  0.188 (0.032) &0.038 (0.007) & 0.193 (0.009) & 0.071 (0.012) & 0.090 (0.008) & 0.068 (0.007) & 0.100\\
\cline{2-11}
& Popular &      0.076 (0.017) & 0.069 (0.008) & 0.268 (0.051) & 0.101 (0.009) & 0.268 (0.045) & 0.080 (0.010) & 0.082 (0.025) & 0.086 (0.008) & 0.129\\
\cline{2-11}
\hline
\multirow{5}{*}{\rotatebox[origin=c]{90}{Prec@1}} & Item-Item &      \textbf{0.941 (0.021)} & \textbf{0.803 (0.034)} &       \textbf{0.917 (0.043)} &\textbf{0.604 (0.056)} & \textbf{0.985 (0.015)} & \textbf{0.765 (0.086)} & \textbf{1.000 (0.000)} & \textbf{0.769 (0.060)} & \textbf{0.848}\\
\cline{2-11}
& ALS &      0.389 (0.020) & 0.374 (0.012) &  0.518 (0.073) &0.177 (0.010) & 0.571 (0.061) & 0.213 (0.030) & 0.451 (0.035) & 0.297 (0.035) & 0.374\\
\cline{2-11}
& BPR &      0.077 (0.031) & 0.037 (0.012) &  0.273 (0.100) &0.017 (0.007) & 0.487 (0.059) & 0.013 (0.004) & 0.212 (0.045) & 0.018 (0.011) & 0.142\\
\cline{2-11}
& Random &      0.108 (0.013) & 0.067 (0.008) &  0.227 (0.052) &0.041 (0.006) & 0.274 (0.023) & 0.067 (0.011) & 0.082 (0.020) & 0.087 (0.010) & 0.119\\
\cline{2-11}
& Popular &      0.054 (0.020) & 0.053 (0.008) &  0.373 (0.075) &0.147 (0.006)& 0.487 (0.059) & 0.112 (0.016) & 0.041 (0.019) & 0.073 (0.006) & 0.1675\\
\cline{2-11}
\hline
\end{tabular}
}
\caption{Evaluation metrics at the artist-level for each algorithm.}
\label{table::artist}
\end{table*}

\section{Experiments} \label{sec:experiments}

For each of these cities, we use the following evaluation procedure:

\begin{algorithm}
\label{euclid}
\begin{algorithmic}[1]
\caption{Evaluation Procedure}\label{alg::eval}
\algnewcommand\algorithmicforeach{\textbf{foreach}}
\algdef{S}[FOR]{ForEach}[1]{\algorithmicforeach\ #1\ \algorithmicdo}


\ForEach {city}
    \ForEach {fold}
        \State construct $X_{train}$ and $X_{eval}$
        \ForEach {algorithm}
            \State train model with $X_{train}$
            \ForEach {playlist $\mathbf{x}^{(p)} \in \mathbf{X}_{eval} $}
                \State split $\mathbf{x}^{(p)}$ into $\mathbf{x}_{non-local}$ and $\mathbf{x}_{local}$ 
                \State use $\mathbf{x}_{non-local}$ with model to 
                \Statex \hspace{70pt} predict $\mathbf{\hat{x}}_{local}$
                \State calculate NDCG, R-Precision, Prec@1 
                \Statex \hspace{70pt} by comparing $\mathbf{x}_{non-local}$ and  
                \Statex \hspace{70pt} $\mathbf{\hat{x}}_{non-local}$ at track- and artist-levels
\EndFor
\EndFor
\EndFor
\EndFor
\end{algorithmic}
\end{algorithm}

\noindent For each city, we partition the local playlists into five equally sized groups and perform five-fold cross-evaluation. That is, we use each group as the evaluation set once and the other four as part of the training set each time.

Using the Implicit Python library\footnote{https://github.com/benfred/implicit}, we calculate Item-Item similarity scores (see Section \ref{sec:iin}) and train both a WRMF model optimized with ALS (see Section \ref{als}) and a matrix factorization model optimized with BPR (see Section \ref{bpr}) using the training set that includes both local and non-local tracks.

For each of the playlists in the evaluation set, we use the non-local tracks $\mathbf{x}_{non-local}$ to generate a ranked list of recommendations based on the score from each of our three algorithms (IIN, ALS, BPR). We have also implemented two baselines: a random baseline which randomly shuffles the local tracks, and a popularity baseline where we rank all of the non-local tracks by their respective popularities. The popularity of a track is estimated as the percentage of playlists that the track appears in from the the training set $X_{train}$.We evaluate each of these five ranked lists on their ability to recommend local music at both the track-level and the artist-level, the latter only looking at the first occurrence of a given artist in the list of track recommendations.

For evaluation metrics, we use the two of the three metrics that were used in the ACM RecSys Challenge 2018 for playlist continuation \cite{chen2018recsys}: Normalized Discounted Cumulative Gain (NDCG) and R-Precision (RPrec).  NDCG evaluates the entire ranking of all $N$ local tracks, weighted such that the top ranked tracks have the greatest importance. The R-Precision for a playlist with $R$ relevant local tracks is the percentage of the $R$ highest scoring local tracks in $\hat{\mathbf{x}}_{local}$ that are present in the ground truth playlist $\mathbf{x}_{local}$. The RecSys Challenge also used a third metric, Clicks, which counted the number of sets of 10 recommended tracks that would be needed before finding the first relevant track. This metric is not appropriate in our setting since we care ranking a much smaller set of tracks (hundreds vs. millions). Instead, we use Precision-at-1 (Prec@1) which measure the accuracy of our top-ranked (i.e. highest scoring) track for each evaluation playlist. By comparision, NDCG reflects the quality of the the entire ranking, RPrec measures the quality of the first few local track recommendations, and Prec@1 is the accuracy of only the top recommendation.     

Our final reported evaluation scores in Tables \ref{table::track} and \ref{table::artist} reflect the averaging of these metrics first over the evaluation playlists in each fold, and then averaged over the five folds. We also report standard error (in parentheses) of each metric over the five folds.

\vspace{-2mm}
\section{Results}\label{sec:results}


As shown in Table \ref{table::track}, the Item-Item Neighborhood model outperforms both baselines (Random, Popularity) and both matrix factorization models (ALS, BPR) in nearly every scenario. The notable exception to this is Chicago, in which the popularity baseline outperformed all other models in all three metrics. This can be explained, however, due to the extremely high sparsity of local tracks. Also, in Chicago's case, while 260 local tracks were found in 84 playlists, the vast majority these playlists contain the same few tracks, preventing the neighborhood model from providing meaningful similarities. Besides this exception, the Item-Item Neighborhood model is consistently the best model, achieving R-Precision and Precision-at-1 scores an order of magnitude better than the other models. 
 
 At the artist-level, shown in Table \ref{table::artist}, the neighborhood model once again performed universally better than the other models and baselines when observing RPrec and Prec@1. For many of the cities, Prec@1 was near perfect, and in the case of Nashville achieved a perfect score of 1. 
 When observing NDCG, the WRMF with ALS model outperforms the Item-Item model for half of the cities. Specifically, these cities have the fewest playlists, artists, and tracks. In most of these cases, the performance of the Item-Item model is comparable to that of ALS.
 
 In both cases, BPR performed significantly worse than expected, even frequently scoring worse than the random model. 
 One potential failing point of this algorithm could be the sparsity of the data. Both of the datasets used to evaluate BPR in \cite{bpr} are much less sparse (less than 99\% sparse), which corresponds to a vastly more dense training and recommendation space than the data used in this experiment.

 In terms of computation overhead time, the Item-Item Neighborhoood model takes the least amount of time to initialize. We conducted our experiment on a 2017 iMac with 16GB of RAM and an Intel Core i5 processor, and the overhead of calculating the Item-Item similarity scores, $s^{(p)}_t$ for every playlist $p$ and track $t$ took about 2 minutes. Training the ALS model took about 20 minutes and training the BPR model took about 2 hours. After training, recommendation time was comparable between all three models.

\vspace{-2mm}
\section{Conclusions}\label{sec:conclusion}

We have presented a novel approach for evaluating local (long-tail) music recommendation. That is, by partitioning a large playlist-track matrix into non-local and local (mostly long-tail) tracks, and considering playlists with one or more these local tracks, we can evaluate how different recommender systems perform on this task. 

Surprisingly, the Item-Item Neighborhood model performs better than the models based on matrix factorization (ALS, BPR) on the task of local music recommendation. 
This opposes the results of \cite{als, bpr}, which show that in general recommendation, ALS and BPR significantly outperform neighborhood models. This may be related to the fact that local (long-tail) music recommendation involves modeling highly sparse data. That is, matrix factorization approaches attempt to optimize parameters to minimize loss of all the ratings. Since the vast majority of the rating are associated with the popular (non-local) tracks in the short-head, these models might not generalize well to local tracks. As a result, a simple Item-Item Neighborhood, which is not susceptible to popularity bias, performs better based on our experiments.   


In future work, we plan to explore modification of matrix factorization models that attempt to mitigate popularity bias. For example, we could re-weight the cost function for the WRMF model (Section \ref{als}) so that the weights are per track rather than per rating.  We also plan to develop and deploy a local music recommendation system to compare performance of recommender system algorithm from the perspective of the end user experience. This requires us to evaluate not only recommendation accuracy but also scalability and robustness in a real-time setting.   

This project is supported by NSF grant IIS-1615679.


\bibliography{ISMIR2019_LocalMusicPlaylists}

\end{document}